\documentclass[english,aps, prb, twocolumn, superscriptaddress,showpacs,floatfix]{revtex4-1}
\usepackage{lmodern}
\usepackage[T1]{fontenc}
\usepackage[latin9]{inputenc}
\usepackage{textcomp}
\usepackage{graphicx}
\usepackage{bm}
\usepackage{verbatim}
\usepackage{booktabs}
\usepackage{amstext}
\usepackage{amsmath}
\usepackage{color}
\usepackage{upgreek}
\usepackage{multirow}
\usepackage{siunitx}
\usepackage{paralist}

\usepackage{printlen}

\makeatletter

\newcommand{\lyxmathsym}[1]{\ifmmode\begingroup\def\b@ld{bold}
  \text{\ifx\math@version\b@ld\bfseries\fi#1}\endgroup\else#1\fi}

\@ifundefined{textcolor}{}
{%
 \definecolor{BLACK}{gray}{0}
 \definecolor{WHITE}{gray}{1}
 \definecolor{RED}{rgb}{1,0,0}
 \definecolor{GREEN}{rgb}{0,1,0}
 \definecolor{BLUE}{rgb}{0,0,1}
 \definecolor{CYAN}{cmyk}{1,0,0,0}
 \definecolor{MAGENTA}{cmyk}{0,1,0,0}
 \definecolor{YELLOW}{cmyk}{0,0,1,0}
 \definecolor{darkgreen}{rgb}{0,0.6,0.4}
\definecolor{azzuri}{rgb}{0.2,0.2,0.7}
 }

\usepackage{prettyref}
\makeatother

\usepackage{babel}

\begin{document}

% Set path to figures
\graphicspath{{.}{Figures/}}

\title{Crystal field states of Tb$^{3+}$ in the pyrochlore spin liquid Tb$_2$Ti$_2$O$_7$ from neutron spectroscopy}

\author{A. J. Princep}
\email[]{a.princep@physics.ox.ac.uk}
\affiliation{Department of Physics, University of Oxford, Clarendon Laboratory,
Parks Road, Oxford, OX1 3PU, United Kingdom}

\author{H. C. Walker}
\affiliation{ISIS Facility, Rutherford Appleton Laboratory, STFC, Chilton, Didcot,
Oxon, OX11 0QX, United Kingdom}

\author{D. T. Adroja}
\affiliation{ISIS Facility, Rutherford Appleton Laboratory, STFC, Chilton, Didcot,
Oxon, OX11 0QX, United Kingdom}

\author{D. Prabhakaran}
\author{A. T. Boothroyd}
\email[]{a.boothroyd@physics.ox.ac.uk}
\affiliation{Department of Physics, University of Oxford, Clarendon Laboratory,
Parks Road, Oxford, OX1 3PU, United Kingdom}

\begin{abstract}
We report time-of-flight neutron scattering measurements of the magnetic spectrum of Tb$^{3+}$ in Tb$_2$Ti$_2$O$_7$. The data, which extend up to 120\,meV and have calibrated intensity, enable us to consolidate and extend previous studies of the single-ion crystal field spectrum. We successfully refine a model for the crystal field potential in Tb$_2$Ti$_2$O$_7$ without relying on data from other rare earth titanate pyrochlores, and we confirm that the ground state is a non-Kramers doublet with predominantly $|\pm 4\rangle$ components. We compare the model critically with earlier models.
\end{abstract}


\maketitle

\section{\label{sec:Introduction} Introduction}

%The \verb|\columnwidth| is \printlength{\columnwidth} which is also and \uselengthunit{mm}\printlength{\columnwidth}.
%The \verb|\textwidth| is \uselengthunit{pt}\printlength{\textwidth} which is also and \uselengthunit{mm}\printlength{\textwidth}.

Among the many magnetically frustrated pyrochlore oxides, Tb$_2$Ti$_2$O$_7$ (TTO) stands out because of its intriguing low temperature state which is thought to be a type of spin liquid.\cite{Gardner-RMP-2010} TTO shows no sign of any conventional symmetry-breaking transition (magnetic or structural) down to temperatures as low as 50\,mK (Refs.~\onlinecite{Gardner-PRL-1999,Gardner-PRB-2003}) despite an antiferromagnetic Curie--Weiss temperature of $-19$\,K (Ref.~\onlinecite{Gingras-PRB-2000}) and predictions of magnetic order at 1--2\,K (Refs.~\onlinecite{denHertog-PRL-2000,Kao-PRB-2003}). There are, however, strong short-range antiferromagnetic correlations in TTO at low temperatures.\cite{Gardner-PRB-2001,Rule-PRL-2006,Guitteny-PRL-2013,Fritsch-PRB-2013}

The spin liquid state in TTO is not fully understood, but a key factor is the low energy part of the crystal-field-split $f$ electron manifold of Tb$^{3+}$, which comprises two doublets separated by about 1.5\,meV (Refs.~\onlinecite{Gardner-PRL-1999,Gingras-PRB-2000}). This splitting is comparable with the exchange and dipolar coupling strengths in TTO. Therefore, although the single-ion ground state is Ising-like, transverse fluctuations of the ground state moment appear once interactions are taken into account, and the cooperative nature of these fluctuations could lead to a quantum spin ice state.\cite{Molavian-PRL-2007,Gingras-RPP-2014} Moreover, the two doublets are also connected by quadrupolar interactions, allowing coupling to phonons. Indeed, the recent observation of hybridization between an acoustic phonon and a crystal field excitation in TTO demonstrates the importance of magnetoelastic interactions in this system.\cite{Guitteny-PRL-2013,Fennell-PRL-2014} So the conditions exist in TTO for enhanced quantum fluctuations and suppressed magnetic ordering via both magnetic and magnetoelastic interactions, and an accurate determination of the crystal field states is required for quantitative modelling.

%The most direct information on the crystal field spectrum of Tb$^{3+}$ in TTO has come from neutron\cite{Gingras-PRB-2000,Gardner-PRB-2001,Mirebeau-PRB-2007,Zhang-PRB-2014} and optical\cite{Lummen-PRB-2008,Maczka-PRB-2008} spectra.
 %measurements of transitions within the crystal-field-split $^7$F$_6$ ground state term of the $4f^{10}$ configuration. Some $^7$F$_6$ to $^7$F$_5$ inter-level transitions have also been observed.\cite{Maczka-PRB-2008}
Several models for the crystal field potential in TTO have been reported based on analyses of neutron and optical spectra combined with thermal and magnetic data.\cite{Gingras-PRB-2000,Gardner-PRB-2001,Mirebeau-PRB-2007,Malkin-PRB-2004,Malkin-JPCM-2010,Klekovkina-JPCS-2011,Klekovkina-Optics-2014,Bertin-JPCM-2012,Zhang-PRB-2014} Unfortunately, there are significant discrepancies between the published sets of crystal field parameters. These are partly caused by the fact that some of the models employ a truncated basis containing only the 13 states of the ground state $^7$F$_6$ manifold, whereas others include all the states in the $f^{8}$ configuration of Tb$^{3+}$. The main difficulty, however, has been that up to now only four transitions within the $J=6$ level have been observed unambiguously. Information from other heavy rare-earth titanates has generally been used to augment the data on TTO so that the six free parameters needed to describe the single-ion crystal-field Hamiltonian can be determined independently.

In a very recent neutron scattering study,\cite{Zhang-PRB-2014} a splitting of one of the peaks was resolved and assumed to originate from two distinct crystal field transitions. This led to a significantly different crystal field potential compared with previous models for TTO. For instance, the wave functions of the ground and first excited doublets are interchanged compared with models informed by data from other rare-earth titanates. The correctness of this assumption has recently been questioned, and the splitting attributed instead to coupling to a phonon\cite{Klekovkina-Optics-2014}.

 The purpose of this paper is to resolve these discrepancies as far as possible. We report neutron scattering measurements of TTO which extend to higher energies than before. We have observed a crystal field transition that was not previously detected, and we have determined the transition intensities on an absolute scale. By fitting the data from TTO alone, using a model that includes intermediate coupling basis states and partial $J$-mixing, we find a crystal field potential that provides a good description of the experimental data. The model described in Ref.~\onlinecite{Zhang-PRB-2014} cannot be reconciled with our data.

%\begin{figure}
%\includegraphics[width=\columnwidth]{LuFe2O4_Bilayer_Structure_v2}
%\caption{\label{fig:LFO_charge_mag_struct}(Color online) Figure caption.}
%\end{figure}

\section{\label{sec:Experimental Details} Experimental Details}

A powder sample of mass 16\,g was prepared by standard solid state synthesis. The sample was found to be single phase and of high quality to within the precision of laboratory X-ray diffraction. Neutron inelastic scattering measurements were performed on the MERLIN time-of-flight spectrometer at the ISIS Facility.\cite{Bewley-Notiziario-2009}  The sample was contained in an aluminium foil packet in the form of an annulus of diameter 40\,mm and height 45\,mm and sealed in an aluminium can containing helium exchange gas. The can was cooled by a closed-cycle refrigerator. Spectra were recorded for approximately 4 hours each with neutrons of incident energy $E_{\rm i}=65$ and 150\,meV at temperatures of $T=7$ and 90\,K. The raw data were corrected for detector efficiency, sample attenuation and time-independent background following standard procedures. Vanadium spectra recorded at the same two incident energies were used to determine the energy resolution and to convert the intensities into units of cross section, mb\,sr$^{-1}$\,meV$^{-1}$\,f.u.$^{-1}$, where f.u. stands for the formula unit of Tb$_2$Ti$_2$O$_7$.

\section{\label{sec:Results} Results}

\begin{figure}
\includegraphics[width=\columnwidth]{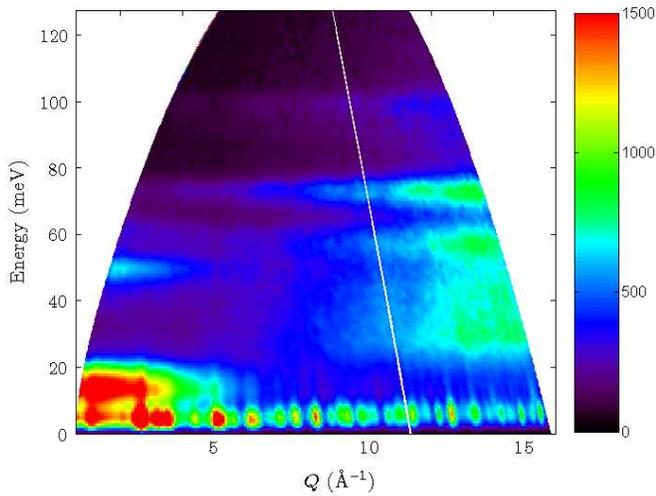}
\caption{\label{fig:E-Q_map}(Color online) Neutron scattering spectrum of polycrystalline Tb$_2$Ti$_2$O$_7$ at $T=7$\,K, measured on MERLIN with $E_{\rm i} = 150$\,meV. The colors represent the value of $E\times S(Q,E)$ in units of mb\, sr$^{-1}$\,f.u.$^{-1}$}.
\end{figure}
Figure~\ref{fig:E-Q_map} provides an overview of the data recorded at $T = 7$\,K with $E_{\rm i}=150$\,meV. The spectrum is presented in the form of a color map of $E\times S(Q,E)$, where $S(Q,E)$ is the scattering intensity as a function of energy $E$ and the magnitude of the scattering vector $Q=|{\bf Q}|$. Multiplication by energy suppresses the strong elastic and low energy inelastic scattering and makes the weaker signals at higher energies more visible.

Previous neutron measurements of the spectrum of TTO revealed magnetic transitions centred at 1.5, 10, 16 and 49\,meV. In Ref.~\onlinecite{Zhang-PRB-2014}, a weak feature near 70\,meV was also attributed to a magnetic transition, and the 16\,meV signal was found to be split into two peaks separated by 2.5\,meV consistent with earlier neutron\cite{Mirebeau-PRB-2007} and Raman\cite{Lummen-PRB-2008} spectra. The 10--16\,meV transitions are clearly visible in Fig.~1 although they not resolved in this particular data set.\cite{footnote-1} The 49\,meV transition is also present and confirmed as magnetic by the characteristic reduction of its intensity with $Q$ due to the magnetic form factor. Above 50\,meV there is a weak feature near 61\,meV which decreases with $Q$, and there are peaks near 72 and 90--100\,meV which increase with $Q$ consistent with scattering from phonons.

\begin{figure}
\includegraphics[width=0.9\columnwidth,trim= 0pt 0pt 0pt 0pt, clip]{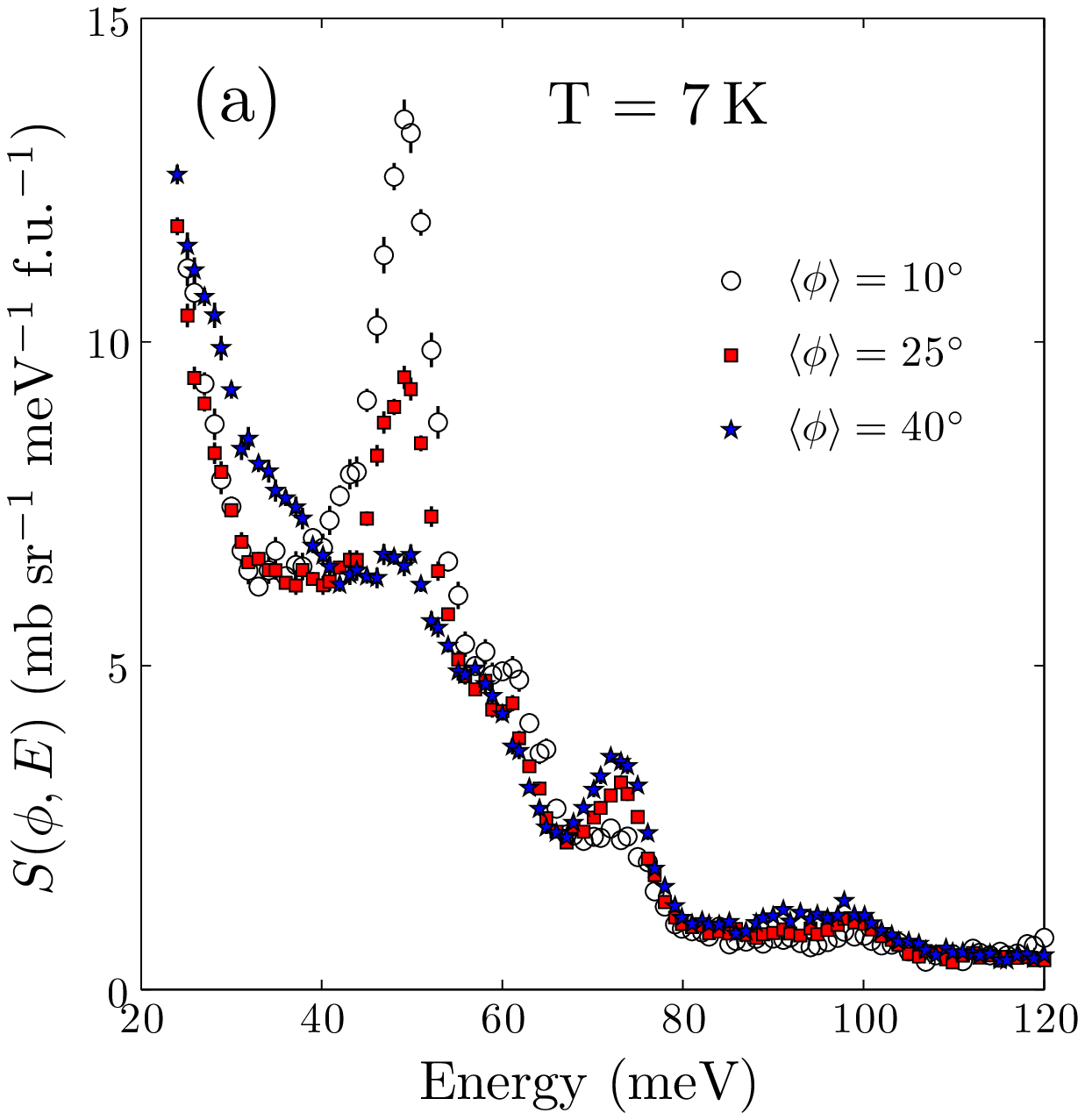}
\includegraphics[width=0.9\columnwidth,trim= 0pt 0pt 0pt 0pt, clip]{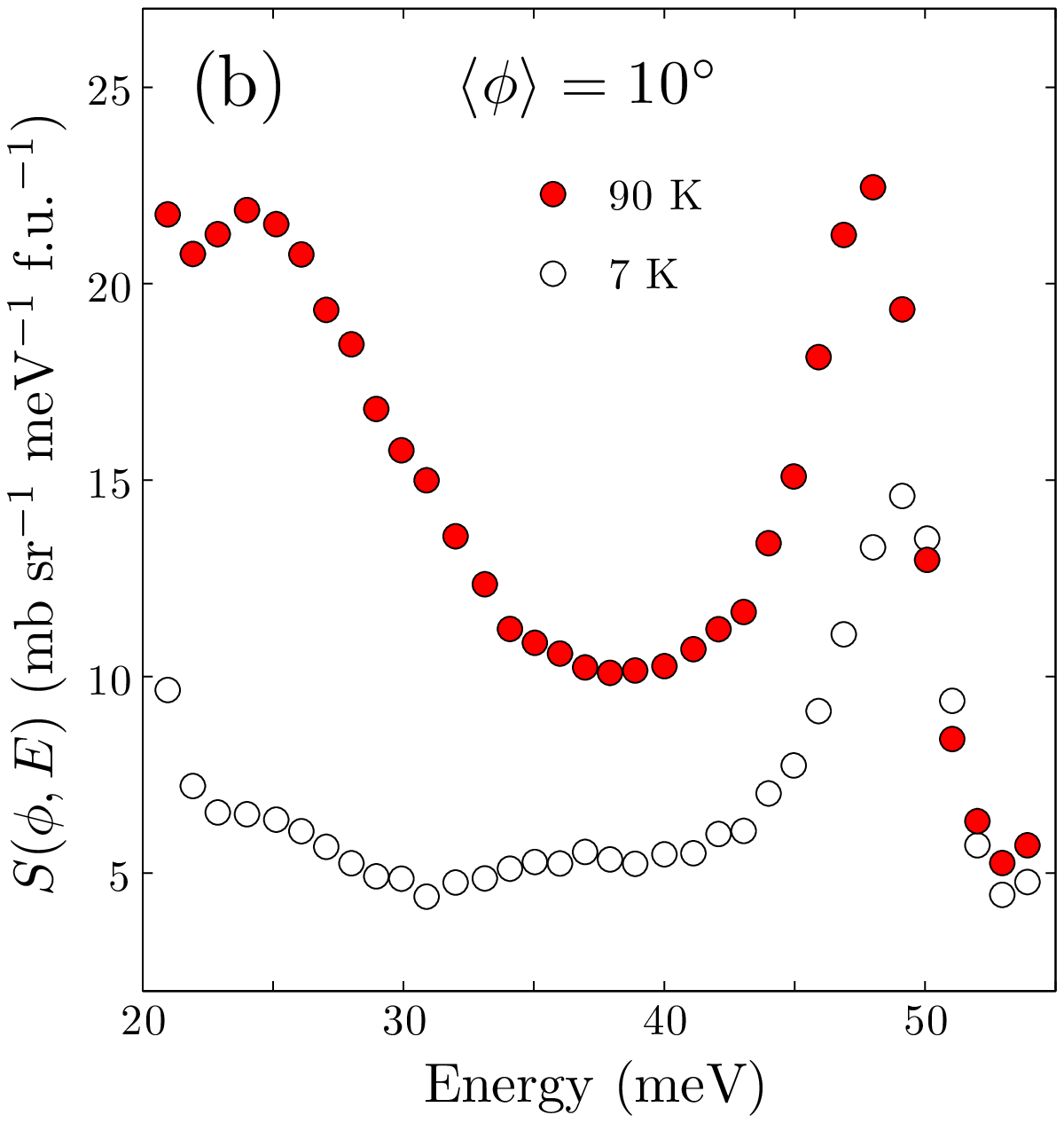}
\caption{\label{fig:spectrum}(Color online) Neutron spectrum $S(\phi,E)$ of polycrystalline Tb$_2$Ti$_2$O$_7$ for fixed scattering angle $\phi$. The corresponding $Q$ is given by $\hbar^2Q^2/2m_{\rm n} = E_{\rm i} + E_{\rm f}-2(E_{\rm i}E_{\rm
f})^{\frac{1}{2}}\cos\phi$, where $E_{\rm f} = E_{\rm i}-E$ and $m_{\rm n}$ is the neutron rest mass. (a) Spectrum at three different average scattering angles recorded with $E_{\rm i} = 150$\,meV at $T=7$\,K, showing magnetic peaks at 49 and 61\,meV and phonon peaks at 72, 91 and 99\,meV. (b) Comparison of the low angle ($\langle \phi \rangle = 10^{\circ}$) spectrum at $T=7$ and 90\,K, recorded with $E_{\rm i} = 65$\,meV.}
\end{figure}
The part of the spectrum with $E>20$\,meV is represented in more quantitative detail in Fig.~\ref{fig:spectrum}. Figure~\ref{fig:spectrum}(a) shows the intensity as a function of energy for three different average scattering angles $\phi$ taken from the $E_{\rm i} = 150$\,meV run at 7\,K.  The large peak at 49\,meV and the small peak at 61\,meV both decrease with increasing $\phi$ (i.e.~increasing $Q$), whereas the peaks at 72, 91 and 99\,meV all increase with angle. This confirms that the latter three peaks are caused by scattering from phonons, to within the sensitivity of the measurement, whereas the 49 and 61\,meV peaks are from magnetic transitions.
%We could find no evidence for any other magnetic peaks in the 7\,K spectrum in the energy range 20 to 125\,meV.

Figure~\ref{fig:spectrum}(b) compares $\phi = 10^{\circ}$ spectra at 7 and 90\,K measured with $E_{\rm i} = 65$\,meV. Two features stand out:

(1) The 49\,meV peak in the 7\,K spectrum is broader than the resolution, with a width (full width at half maximum) of 4.9\,meV as compared with the instrumental resolution of about 1.5\,meV at this energy. The peak is also asymmetric, with a low energy tail that is not wholly accounted for by the asymmetry of the resolution function. On warming to 90\,K, the peak increases in intensity and shifts down in energy to 47.5\,meV. Transitions from the thermally populated 1.5\,meV first excited crystal field level account for some of the peak width at 7\,K but not all, and there are no sharp phonon peaks at this energy (see Fig.~\ref{fig:E-Q_map}). These observations suggest that there exists an additional crystal field level at about $48$\,meV which is strongly connected to the 1.5\,meV level.  The peak positions and widths at 7\,K and 90\,K would then be explained by transitions from the ground state and 1.5\,meV level to levels at 48 and 49\,meV.

(2) In the range 20--45\,meV there is considerable additional scattering at 90\,K relative to 7\,K. The 90\,K scattering does not display any particularly prominent features, but a small peak has grown at about 24\,meV and there is a weak shoulder near 29\,meV which coincides with a minimum in the 7\,K spectrum. These temperature-induced features are not accounted for by the thermal population of phonons (the temperature factor for phonon scattering $[1-\exp(-E/k_{\rm B}T)]^{-1}$ increases by only 4\% at 24\,meV on warming from 7 to 90\,K). The additional scattering must therefore derive from transitions out of thermally excited magnetic levels, and since the separation between the peaks near 16 and 49\,meV is significantly greater than 25\,meV there must exist a hitherto undetected level between these two. Indeed, thermally excited transitions from the 10 and 16\,meV levels to a level near 39\,meV would give rise to enhanced intensity around 24 and 29\,meV, as observed.

In short, we have observed a magnetic peak at 61\,meV which corresponds to a crystal field level not found in previous studies, and we have indirect evidence for additional levels near 39\,meV and 48\,meV.

To constrain the crystal field model for TTO as tightly as possible we shall also consider the transition intensities. We determined the integrated intensities of the magnetic peaks in the low angle spectrum at 7\,K by fitting an asymmetric pseudo-Voigt line shape to the peaks. The parameters of the pseudo-Voigt function were determined according to an empirical implementation of the analytic line shape which contains a contribution from the velocity selection via the chopper,\cite{Windsor} and an additional component due to the pulse width. The nonmagnetic (phonon) background was estimated from the high angle part of the spectrum, which was scaled to match the high energy transfer part of the low angle spectrum. A list of observed energy levels and transition intensities at 7\,K is given in Table~\ref{table1}. For ease of comparison, the intensities have been extrapolated to zero $Q$ via the $Q$ dependence of the magnetic dipole form factor of Tb$^{3+}$.
\begin{table}
\caption{\label{table1} Observed and calculated crystal-field transition energies and integrated intensities from the spectrum of Tb$_2$Ti$_2$O$_7$ measured at 7\,K. Numbers in parentheses are experimental errors in the last digit. Observed intensities have been corrected for the dipole form factor of Tb$^{3+}$ so that both $I_{\rm obs}$ and $I_{\rm calc}$ are for zero $Q$. The $I_{\rm calc}$ values include transitions from both 0.0 and 1.4\,meV levels with their respective thermal populations at 7\,K. The best-fit crystal-field parameters used for the calculations are: $B^2_0 = 55.3$, $B^4_0 = 370.4$, $B^4_3 = 128.0$, $B^6_0 = 114.3$, $B^6_3 = -114.3$, $B^6_6 = 120.3$\,meV.}
\begin{ruledtabular}
\begin{tabular}{ccccc}
\textrm{Level}&
\textrm{$E_{\rm obs}$}&
\textrm{$E_{\rm calc}$}&
\textrm{$I_{\rm obs}$}(Q=0)&
\textrm{$I_{\rm calc}(Q=0)$}\\
 & (meV) & (meV) & (mb\,sr$^{-1}$\,f.u.$^{-1}$) & (mb\,sr$^{-1}$\,f.u.$^{-1}$)\\
\colrule
$\Gamma_3^+$ & 0.0 & 0.0 & $-$ & 2918\\[2pt]
$\Gamma_3^+$ & 1.4(4)\footnote{The first excited state is dispersive at low temperature with a band width approaching 1\,meV.\cite{Gardner-PRL-1999,Guitteny-PRL-2013,Mirebeau-PRB-2007,Zhang-PRB-2014,Kanada-JPSJ-1999}} & 1.5 & $-$ & 2744\\[2pt]
$\Gamma_2^+$  & 10.2(4) & 10.3  & 1740(60) & 1836\\[2pt]
$\Gamma_1^+$  & 16.0(5)\footnote{Comprises two peaks separated by 2.5\,meV.\cite{Mirebeau-PRB-2007,Zhang-PRB-2014}} & 16.1 & 990(100) & 1010 \\[2pt]
$\Gamma_3^+$  & 39(2) & 39.0 & $\rceil  \hspace{35pt}$  & 31\\[2pt]
$\Gamma_2^+$  & 48(1) & 48.2 & | 230(35)  & 62\\[2pt]
$\Gamma_1^+$  & 49(1) & 48.8& $\rfloor \hspace{35pt}$ & 184\\[2pt]
$\Gamma_3^+$  & 61(1) & 60.8 & 44(6) & 52\\[2pt]
$\Gamma_1^+$  & $-$ & 71.0 & $-$ & 8\\
\end{tabular}
\end{ruledtabular}
\end{table}

The crystal field at the Tb$^{3+}$ site in TTO has point symmetry $\overline{3}m$ ($D_{3d}$) and is described by the Hamiltonian
\begin{eqnarray}
\mathcal{H}_{\rm CF} = & B^2_0C^2_0 + B^4_0C^4_0 + B^4_3(C^4_{-3}-C^4_3) + B^6_0C^6_0 \nonumber \\
&+ B^6_3(C^6_{-3}-C^6_3)+ B^6_6(C^6_{-6}+C^6_6),
\label{eq:Hamiltonian}
\end{eqnarray}
where $B_q^k$ are the crystal field parameters and $C_q^k$ are Wybourne tensor operators.\cite{Wybournebook} Diagonalisation of $\mathcal{H}_{\rm CF}$ was performed in the intermediate coupling scheme with the program SPECTRE.\cite{SPECTRE}
%The various electrostatic and spin--orbit parameters of the free-ion Hamiltonian were taken from Carnall {\it et al.}\cite{Carnall-JCP-1989}
To speed up the calculation, the complete basis of the $f^8$ configuration of Tb$^{3+}$ (3003 states) was truncated to the lowest 110 states (the complete $^7F$ term, plus the states $^5D_4$, $^5D_3$, $^5G_6$, $^5L_{10}$ and $^5G_5$) extending to 3.5\,eV above the ground state.  The diagonalisation of $\mathcal{H}_{\rm CF}$ included $J$-mixing within the truncated basis.

The neutron spectrum of single-ion magnetic transitions is given by\cite{neutronbook}
\begin{eqnarray}
S({\bf Q},E) & = & \left(\frac{\gamma r_0}{2}\right)^{\hspace{-2pt}2}{\rm e}^{-2W}\sum_{i}p_{i}\sum_{j} \nonumber \\
& & \times |\langle\Gamma_j|\,{\bf M}_{\perp}({\bf Q})\,|\Gamma_i\rangle|^2\delta(E_{j} - E_{i}-E),
\label{eq:S(Q,w)_CF_transitions}
\end{eqnarray}
where $(\gamma r_0/2)^2 = 72.7$\,mb. The first summation is over the initial states $\Gamma_i$ with thermal population $p_i$, and the second summation is over the final states $\Gamma_j$. The Debye--Waller factor ${\rm e}^{-2W}$ is taken to be unity at the low temperatures of the measurements. We assume the dipole approximation, in which case $ {\bf M}_{\perp}({\bf Q})$ can be replaced by $-f(Q)g_J{\bf J}_{\perp}$
%\begin{equation}
% {\bf M}_{\perp}({\bf Q}) = -f(Q)g_J{\bf J}_{\perp},
% \label{M_perp_dipole_approx}
%\end{equation}
where $f(Q)$ is the dipole form factor, $g_J$ is the Land\'{e} $g$-factor and ${\bf J}_{\perp}$ is the component of the total angular momentum perpendicular to $\bf Q$. Calculated intensities are powder-averaged for comparison with the data.

The six crystal field parameters in Eq.~(\ref{eq:Hamiltonian}) were refined by a weighted least-squares fitting algorithm against the experimental data given in Table~\ref{table1}, with the intensities expressed relative to the 10.2\,meV peak. The squares of the experimental uncertainties were used as reciprocal weights.  The crystal field parameters determined for Ho$_2$Ti$_2$O$_7$ (Ref.~\onlinecite{Rosenkranz-JAP-2000}) were used as starting parameters for the fit. The procedure converged to an excellent fit with $\chi^2 = 0.95$, where $\chi^2$ is the standard normalised goodness-of-fit statistic. Fits were also performed with different sets of starting parameters but no other acceptable distinct solutions were found. In particular, fits starting from the parameters found by Zhang {\it et al.}\cite{Zhang-PRB-2014} did not converge.

\begin{figure*}
\includegraphics[width=1.8\columnwidth,trim= 0pt 0pt 0pt 0pt, clip]{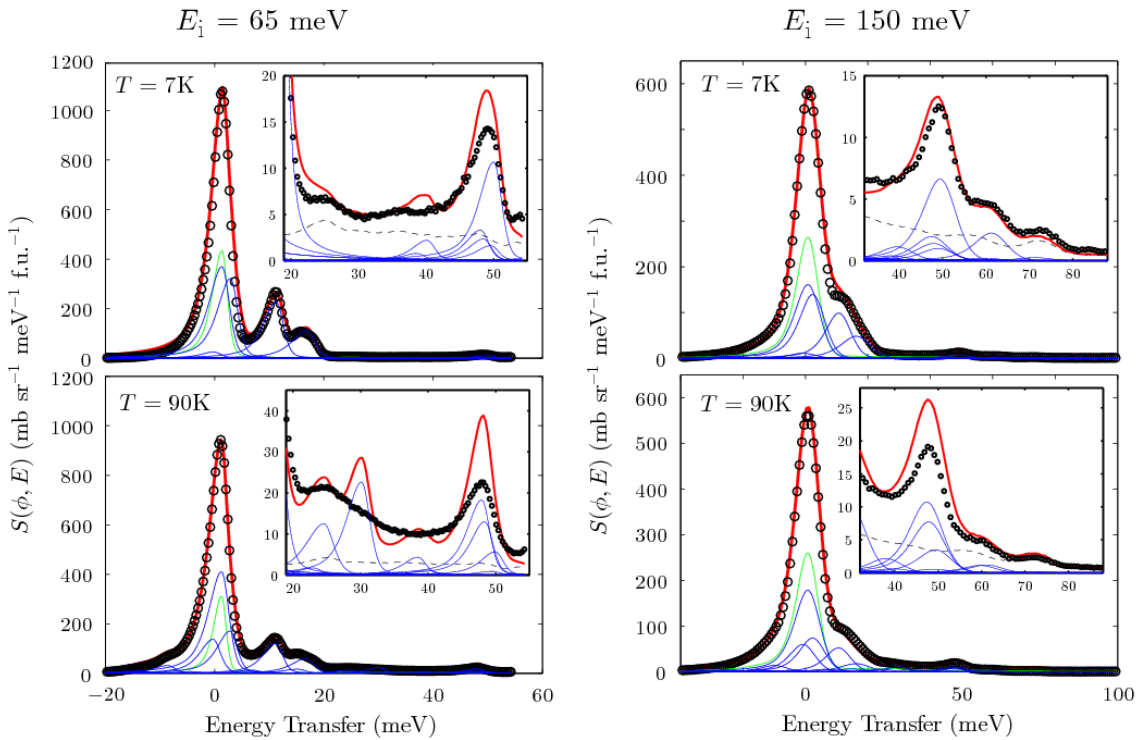}
\caption{\label{fig:fit}(Color online) Comparison of the observed and calculated spectrum of polycrystalline Tb$_2$Ti$_2$O$_7$. The heavy solid (red) lines are calculated from Eq.~(\ref{eq:S(Q,w)_CF_transitions}) with eigenstates derived from the crystal field parameters given in Table~\ref{table1}, and the individual transitions are shown as thin (blue) lines. The broken (blue) lines are the estimated non-magnetic background, and the fitted elastic peaks centred on $E=0$ are shown as pale (green) lines.}
\end{figure*}

The best-fit parameters are given in Table~\ref{table1} together with the calculated energy levels and intensities at 7\,K. The calculated values agree very well with the observations, including the absolute intensities which have a systematic error of 5--10\% from uncertainties in the vanadium calibration and in the corrections for attenuation and the $Q$ dependence of the magnetic form factor. Figure~\ref{fig:fit} compares the predictions of the best-fit model with the spectra measured at 7\,K and 90\,K. Pseudo-Voigt line shapes have been used to model the resolution function. The intensities are calculated in absolute units from Eq.~(\ref{eq:S(Q,w)_CF_transitions}) and have not been scaled to fit the data. Overall, the agreement is very good. The main discrepancy is that the measured spectrum at 90\,K is less structured above 20\,meV than the calculated spectrum, which suggests that the crystal field levels in this energy range broaden significantly with temperature.
\begin{table}
\caption{\label{table2} Comparison of crystal field parameters $B^k_q$ (in meV) for Tb$_2$Ti$_2$O$_7$ from different analyses. The parameters are defined for the Hamiltonian in Eq.~(\ref{eq:Hamiltonian}). The effective Hamiltonian in terms of Stevens operators $O^q_k$,\cite{Hutchings} which applies for a basis comprising a single $J$ level in the $LS$ coupling scheme, is $\mathcal{H}_{\rm CF} = \sum_{k,q}\theta_k D^q_k O^q_k$.  The $\theta_k$ are reduced matrix elements which for Tb$^{3+}$ ($J=6$) are $\theta_2 = -1/99$, $\theta_4 = 2/16335$ and $\theta_6 = -1/891891$. The parameters $D^q_k$ and $B^k_q$ are related by $D^q_k = \lambda^q_kB^k_q$, where $\lambda^0_2 = 1/2$, $\lambda^0_4 = 1/8$, $\lambda^3_4 = \sqrt{35}/2$, $\lambda^0_6 = 1/16$, $\lambda^3_6 = \sqrt{105}/8$ and $\lambda^6_6 = \sqrt{231}/16$. }
\begin{ruledtabular}
\begin{tabular}{ccccccc}
\textrm{Ref.} & $B^2_0$ & $B^4_0$ & $B^4_3$ & $B^6_0$ & $B^6_3$ & $B^6_6$\\
\colrule
\onlinecite{Gingras-PRB-2000} & 53.6 & 318 & 146 & 149 & $-143$ & 67.6\\
\onlinecite{Mirebeau-PRB-2007} & 60.9 & 291 & 103 & 96.6 & $-59.9$ & 97.5\\
\onlinecite{Klekovkina-Optics-2014}\footnote{The model in Ref.~\onlinecite{Klekovkina-Optics-2014} is a refinement of the models presented in Refs.~\onlinecite{Malkin-PRB-2004,Malkin-JPCM-2010,Klekovkina-JPCS-2011}} & 56.0 & 329 & 95 & 107 & $-77.4$ & 109\\
\onlinecite{Bertin-JPCM-2012} & 67.3 & 320 & 119 & 113 & $-90.5$ & 101\\
\onlinecite{Zhang-PRB-2014} & 144 & 268 & 162 & 171 & 349 & 799\\
\onlinecite{Rosenkranz-JAP-2000}\footnote{Parameters for Ho$_2$Ti$_2$O$_7$ scaled with the point-charge relation\cite{Hutchings} $B^k_q({\rm Tb})/B^k_q({\rm Ho}) = \langle r^k \rangle_{\rm Tb}/\langle r^k\rangle_{\rm Ho}$, where $\langle r^k \rangle$ is the $k^{\rm th}$ radial moment of the 4$f$ electron distribution.} & 75.3 & 329 & 100 & 111 & $-79.6$ & 130\\
\textrm{This work} & 55.3 & 370 & 128 & 114 & $-114$ & 120\\
\end{tabular}
\end{ruledtabular}
\end{table}

\section{\label{sec:Discussion} Discussion}

We have presented here a single-ion model based solely on measurements on Tb$_2$Ti$_2$O$_7$ that successfully describes the general features of the observed magnetic spectrum of Tb$^{3+}$ in Tb$_2$Ti$_2$O$_7$.  The crystal field parameters of our model are compared with previously published sets [converted to the Wybourne tensor parameters of Hamiltonian~(\ref{eq:Hamiltonian})] in Table~\ref{table2}. Our parameters have similar magnitudes and signs to those obtained for TTO in Refs.~\onlinecite{Gingras-PRB-2000} and \onlinecite{Mirebeau-PRB-2007,Malkin-PRB-2004,Malkin-JPCM-2010,Klekovkina-JPCS-2011,Klekovkina-Optics-2014,Bertin-JPCM-2012} (which however relied on spectroscopic data on other rare earth titanates). The crystal field parameters are also similar to those of Ho$_2$Ti$_2$O$_7$ (Ref.~\onlinecite{Rosenkranz-JAP-2000}) and Pr$_2$Sn$_2$O$_7$ (Ref.~\onlinecite{Princep-PRB-2013}) after taking into account the differences in the radial moments of the respective $4f$ orbitals. The lack of significant variation shown in the crystal field potential across different systems implies that the local structure and bonding is similar for different pyrochlore oxides.

As deduced many years ago,\cite{Gingras-PRB-2000} the low energy part of the spectrum at low temperature is composed of two non-Kramers doublets separated by approximately 1.5\,meV. From our model, the largest components of the ground state and first excited doublet wave functions are found to be
\begin{eqnarray}
\begin{split}
\Gamma_3^+ (0\,{\rm meV})  = \ &  0.968|^7F_6,\pm 4 \rangle \mp  0.089|^7F_6,\pm 1\rangle \nonumber \\ \ & + 0.110|^7F_6,\mp 2\rangle \pm 0.181 |^7F_6,\mp 5\rangle \nonumber\\
 & - 0.083|^7F_4, \pm 4\rangle     \nonumber\\[5pt]
 \Gamma_3^+ (1.5\,{\rm meV})  =  \ &  0.952|^7F_6,\pm 5 \rangle \mp 0.176|^7F_6,\pm 2 \rangle \\ & + 0.061|^7F_6,\mp 1 \rangle \pm 0.194|^7F_6,\mp 4 \rangle \\ & \pm 0.134|^7F_5,\pm 5 \rangle.  \nonumber
 \end{split}
\label{eq:wavefunctions}
\end{eqnarray}
The upper and lower signs give the two components of the doublet, and $|^{2S+1}L_J,m_J\rangle$ identifies the spectroscopic term and $m_J$ value for each component. The dominant components in the wave functions are from the Hund's rule ground state term $|^7F_6\rangle$, as expected, but there is a non-negligible admixture of the $|^7F_5\rangle$ and $|^7F_4\rangle$ states which in the free ion lie 250 and 410\,meV, respectively, above the ground state. For comparison, with the same crystal field model but working in the pure $|^7F_6\rangle$ basis with the Stevens operator form of $\mathcal{H}_{\rm CF}$ instead of Eq.~(\ref{eq:Hamiltonian}) we find that the overall splitting increases from 71\,meV to 81\,meV. A list of the wave functions for all the levels within the $|^7F_6\rangle$ manifold is given in the Appendix.

Our results are consistent with most previous studies, but are in contrast to the work of Zhang {\it et al.}\cite{Zhang-PRB-2014} who found essentially the same two lowest doublets as above but in the reverse order, i.e.~with $|^7F_6,\pm 5\rangle$ as the dominant components of the ground state doublet. This difference arises because Zhang {\it et al.} assumed in their analysis that the splitting of the 16\,meV peak was due to two distinct crystal field transitions. Although the model of Zhang {\it et al.} fits their own data well, our measurements reveal a number of problems with it. Firstly, the magnetic peak we observe at 61\,meV was not identified as a crystal field transition by Zhang {\it et al.}, and therefore not used to constrain the model. Second, their model predicts a level at 101\,meV with a very large cross section for transitions to it from the ground state doublet: 440\,mb\,sr$^{-1}$\,f.u.$^{-1}$ at zero $Q$. This prediction is inconsistent with our data since such an intense transition would produce a peak centred at 101\,meV in Figs.~\ref{fig:E-Q_map} and \ref{fig:spectrum}(a) which, after correction for the Tb$^{3+}$ form factor, would be of roughly the same size as the peak at 49\,meV. Third, the model of Zhang {\it et al.} predicts a strong transition from the first excited doublet to a level near 72\,meV. The zero-$Q$ cross section for this transition, which should appear as a thermally excited peak near 70\,meV, is about 270\,mb\,sr$^{-1}$\,f.u.$^{-1}$ at $T=90$\,K, inconsistent with the 90\,K spectrum shown in the lower right panel of Fig.~\ref{fig:fit}. Finally, the significant intensity from thermally excited transitions observed between 20 and 40\,meV --- see Fig.~\ref{fig:spectrum}(b) --- is not reproduced.

\begin{table*}[!]
\caption{\label{table3} Eigenfunctions of the best-fit single-ion crystal field model for Tb$^{3+}$ in Tb$_2$Ti$_2$O$_7$. The crystal field Hamiltonian is given in Eq.~(\ref{eq:Hamiltonian}), and the values of the best-fit crystal field parameters are given in the caption of Table~\ref{table1}. The list below is of the energy levels within the $|^7F_6\rangle$ manifold and gives only the coefficients of $|m_J\rangle$ belonging to $|^7F_6\rangle$. A blank entry means a zero coefficient.}
\begin{ruledtabular}
\begin{tabular}{rrrrrrrrrrrrrrr}
 & & & & & \multicolumn{3}{c}{Energy (meV)} & & & & & & \\[2pt]
 $m_J$ & 0.0 & 0.0 & 1.5 & 1.5 & 10.3 & 16.1 & 39.0 & 39.0 & 48.2 & 48.8 & 60.8 & 60.8 & 71.0 \\[2pt]
 \colrule
$-6$ & & & & & 0.145 & $-0.179$ & & & 0.687 & 0.678 & & & $-0.033$ \\
$-5$ & 0.181 & & 0.952 & & & & $-0.182$ & & & & 0.033 & &  \\
$-4$ & & 0.968 & & 0.194 & & & & $-0.071$ & & & & $-0.096$ & \\
$-3$ & & & & & 0.686 & $-0.675$ & & & $-0.134$ & $-0.173$ & & & $-0.066$ \\
$-2$ & 0.110 & & 0.176 & &  & & 0.956 & & & & 0.057 & &  \\
$-1$ & & 0.089 & & 0.061 & & & & $-0.058$ & & & & 0.974 &  \\
$0$ & & & & & & $-0.105$ & & & & 0.019 & & & 0.982 \\
$+1$ & $-0.089$ & & 0.061 & & & & 0.058 & & & & 0.974 & &  \\
$+2$ & & 0.110 & & $-0.176$ & & & & 0.956 & & & & 0.057 &  \\
$+3$ & & & & & 0.686 & 0.675 & & & $-0.134$& 0.173 & & & 0.066 \\
$+4$ & 0.968 & & $-0.194$ & & & & $-0.071$ & & & & 0.096 & &  \\
$+5$ & & $-0.181$ & & 0.952& & & & 0.182 & & & & $-0.033$ &  \\
$+6$ & & & & & $-0.145$ & $-0.179$ & & & $-0.687$ & 0.678 & & & $-0.033$ \\

\end{tabular}
\end{ruledtabular}
\end{table*}

The success of our model, which assumes that the split peak at 16\,meV corresponds to a single crystal field level, raises questions about the nature of this excitation.  One possibility is that the splitting could be the result of disorder, e.g.~Tb/Ti site mixing, which could produce two slightly different local environments for the Tb site. However, the fact that no splitting is detectable in any of the other transitions, especially the $\Gamma_3$ doublets whose degeneracy is lifted once the $\overline{3}m$ symmetry is broken, suggests that there is only one Tb$^{3+}$ environment.  Further, the splitting is too large to be explained by the same two-ion magnetic coupling model which accounts for the dispersion of the first excited doublet.\cite{Kao-PRB-2003} Klekovkina and Malkin have suggested that the peak splitting is caused by coupling to a phonon with an energy near 16\,meV at the zone centre.\cite{Klekovkina-Optics-2014} Indeed, hybridization of acoustic phonons with the 1.5\,meV and 10\,meV crystal field levels has recently been observed in neutron spectra of TTO,\cite{Guitteny-PRL-2013,Fennell-PRL-2014} and strong thermally-induced phonon anomalies are observed and attributed to magneto-phonon coupling.\cite{Maczka-PRB-2008} The existence of magnetoelastic modes was suggested as an explanation for why frustration is not relieved in TTO by any conventional symmetry-breaking transitions at low temperatures.\cite{Guitteny-PRL-2013,Fennell-PRL-2014} Although our data cannot shed much light on this interesting question, we mention that the crystal field states in TTO do have strong quadrupole moments so a perturbation to the single-ion states involving a coupling via orbitals is plausible. Measurements on single crystals are needed to identify any coupling to specific phonon modes and to see whether the size of the splitting varies throughout the Brillouin zone.
%{\color{red}it is worth mentioning that there is a phonon peak centred on 17\,meV in our spectra at high $Q$ (where the magnetic form factor of Tb$^{3+}$ is negligible).   Show this as an insert to one of the figures, or a separate figure perhaps? We should calculate the matrix elements of the quadrupole operators connecting the ground state to the 16\,meV level. If one or more of the quadrupole operators has a strong matrix element then this would support the idea of coupling to vibrations of the lattice.}

As far as the physical properties of TTO are concerned, the most important result from this study is that the ground state is dominated by the components $|^7F_6,\pm 4\rangle$.  This means that at temperatures where the first excited doublet has negligible population TTO has strong Ising-like anisotropy with the moments confined to the quantization axis, i.e.~to the local $\langle 111 \rangle$ directions. For the ground state doublet, we calculate that the components of the zero-field spectroscopic $g$-tensor parallel and perpendicular to the quantization axis are $g_{\parallel}=10.7$ and $g_{\perp}=0$ ($g_{\perp}=0.07$ in a field of 1 Tesla), consistent with previous estimates.\cite{Bertin-JPCM-2012}
%{\color{red} The zero-temperature magnetic moment induced by a field of 1\,Tesla is 5.64\,$\mu_{\rm B}$. {\color{red} However, this will be sensitive to coupling/correlation effects, so it might be better to compare at higher fields and temperatures. Low temperature magnetization: Lhotel {\it et al.} PRB {\bf 86}, 020410 (2012); Legl {\it et al.} PRL {\bf 109}, 047201 (2012). }

\section{\label{sec:Conclusion} Conclusion}

We have successfully refined a model for the crystal field in Tb$_2$Ti$_2$O$_7$ against the magnetic spectrum measured by neutron spectroscopy. The model is in very good quantitative agreement with the experimental spectrum, including the absolute intensity, and confirms that the single-ion magnetic properties at low temperature are controlled by two non-Kramers doublets separated by about 1.5\,meV. The wave functions of these states are approximately $|m_J\rangle = |\pm 4\rangle$ and $|\pm 5\rangle$, respectively. Our analysis is more tightly constrained by experiment than has hitherto been possible, and resolves an uncertainty in the literature about the composition of the ground state. The splitting of the 16\,meV peak remains unexplained, and needs further investigation.

\begin{acknowledgments}
We are grateful to Michel Gingras and Bruce Gaulin for helpful comments. This work was supported by the UK Engineering \& Physical Sciences Research Council.\\[5pt]
\end{acknowledgments}

\appendix
\section{\label{app:Wavefunctions}Single-ion eigenfunctions of Tb$_2$Ti$_2$O$_7$}

The eigenfunctions for the best-fit model are given in Table~\ref{table3}.

%\bibliographystyle{apsrev4-1}
%\bibliography{LuFe2O4_bib}

\begin{references}
\bibitem{footnote-1} The 10 and 16\,meV peaks are resolved in our $E_{\rm i} = 65$\,meV spectrum --- see Fig.~\ref{fig:fit}.  The energy resolution of this run is not sufficient to resolve the splitting of the 16\,meV peak observed in other samples, but the 16\,meV peak is broader than the 10\,meV peak and has two shoulders consistent with the 2.5\,meV splitting reported previously. It is likely, therefore, that the splitting of this peak is present in the spectrum of our sample too.
\end{references}

\end{document}